\documentclass[aps,prd,reprint,superscriptaddress,floatfix,showpacs]{revtex4-2}
\pdfoutput=1
\usepackage{graphicx}
\usepackage{cleveref}
\usepackage[english]{babel}
\usepackage{placeins}
\usepackage{longtable}
\usepackage{dcolumn}
\usepackage{xcolor}
\usepackage{siunitx}

\begin{document}

\title{\boldmath Search for the reaction $e^{+}e^{-} \rightarrow \chi _{cJ} \pi ^{+}\pi ^{-} $ and a charmonium-like structure decaying to $\chi _{cJ} \pi ^{\pm}$ between 4.18 and 4.60~GeV}

\author{
\begin{small}
M.~Ablikim$^{1}$, M.~N.~Achasov$^{10,c}$, P.~Adlarson$^{64}$,
S. ~Ahmed$^{15}$, M.~Albrecht$^{4}$, A.~Amoroso$^{63A,63C}$,
Q.~An$^{60,48}$, ~Anita$^{21}$, Y.~Bai$^{47}$, O.~Bakina$^{29}$
% ...
\\
\vspace{0.2cm}
(BESIII Collaboration)\\
\vspace{0.2cm} {\it
$^{1}$ Institute of High Energy Physics, Beijing 100049, People's Republic of China\\
% ...
$^{4}$ Bochum Ruhr-University, D-44780 Bochum, Germany\\
% ...
$^{10}$ G.I. Budker Institute of Nuclear Physics SB RAS (BINP), Novosibirsk 630090, Russia\\
% ...
$^{15}$ Helmholtz Institute Mainz, Johann-Joachim-Becher-Weg 45, D-55099 Mainz, Germany\\
% ...
$^{21}$ Indian Institute of Technology Madras, Chennai 600036, India\\
% ...
$^{29}$ Joint Institute for Nuclear Research, 141980 Dubna, Moscow region, Russia\\
% ...
$^{47}$ Southeast University, Nanjing 211100, People's Republic of China\\
$^{48}$ State Key Laboratory of Particle Detection and Electronics, Beijing 100049, Hefei 230026, People's Republic of China\\
% ...
$^{60}$ University of Science and Technology of China, Hefei 230026, People's Republic of China\\
% ...
$^{63}$ (A)University of Turin, I-10125, Turin, Italy; (B)University of Eastern Piedmont, I-15121, Alessandria, Italy; (C)INFN, I-10125, Turin, Italy\\
$^{64}$ Uppsala University, Box 516, SE-75120 Uppsala, Sweden\\
% ...
$^{c}$ Also at the Novosibirsk State University, Novosibirsk, 630090, Russia\\
% ...
}
\vspace{0.4cm}
\end{small}
}

\noaffiliation{}

\begin{abstract}
We search for the process $e^{+}e^{-}\rightarrow \chi _{cJ} \pi ^{+}\pi ^{-} $ ($J=0,1,2$) and for a charged charmonium-like state in the $\chi _{cJ} \pi ^{\pm}$ subsystem.
The search uses data sets collected with the BESIII detector at the BEPCII storage ring at center-of-mass energies between 4.18~GeV and 4.60~GeV.
No significant $\chi _{cJ} \pi ^{+}\pi ^{-} $ signals are observed at any center-of-mass energy, and thus upper limits are provided which also 
serve as limits for a possible charmonium-like structure in the invariant $\chi _{cJ} \pi ^{\pm} $ mass.
\end{abstract}



\maketitle

\section{Introduction}
In the past decade, the discovery of new and exotic resonances has opened up exciting possibilities for further study of quantum chromodynamics in the charmonium and bottomonium energy regions~\cite{Godfrey::2008,Brambilla::2011,Brambilla::2014}.
One important resonance is the $Y(4260)$, which was observed by the BaBar collaboration in the initial state radiation (ISR) process $e^{+}e^{-} \rightarrow \gamma_{\text{ISR}}\,J/\psi\,\pi^{+}\pi^{-}$~\cite{BABAR::2005,BABAR::2012} and was confirmed by several other collaborations, such as CLEO~\cite{He::2006}, 
Belle~\cite{Yuan::2007,Liu::2013} and BESIII~\cite{Ablikim::2013}. Despite lying above several open-charm thresholds (starting at 3.73\,GeV/$c^{2}$), the $Y(4260)$ state, with quantum number $J^{PC} = 1^{--}$, unconventionally couples much more strongly to the 
final state $J/\psi\,\pi^{+}\pi^{-}$~\cite{Mo::2006} rather than to open-charm final states. This unexpected behavior has stimulated much interest in the hadron-spectroscopy community.
\par
In 2008, the Belle collaboration, studying the decay $\bar{B}^{0} \rightarrow K^{-}\pi ^{+} \chi _{c1}$, observed two charged charmonium-like structures in the $\chi _{c1} \pi ^{\pm}$ sub-system with a statistical significance of $5\sigma$.  These structures were denoted as the  
$Z_{c}(4050)^{\pm}$ and the $Z_{c}(4250)^{\pm}$, with masses of $4051 \pm 14 ^{+20}_{-41}$\,MeV/$c^{2}$
and $4248 ^{+44+180}_{-29-35}$\,MeV/$c^{2}$, respectively, and corresponding widths of $82^{+21+47}_{-17-22}$ and
$177^{+54+316}_{-39-61}$\,MeV~\cite{Mizuk:2008aa}. 
This observation was not confirmed by BaBar, who set $90\%$ confidence level on the presence of these intermediate states~\cite{Lees:2012aa}. 
The first charged charmonium-structure to be found was the $Z(4430)^{\pm}$ decaying to $\psi(2S)\pi^{\pm}$, observed by Belle~\cite{PhysRevLett.100.142001}, whose resonance nature was established by the LHCb collaboration~\cite{Aaij:2014jqa}. The presence of an electric charge indicates a possible internal structure of at least four quarks.
\par
In order to gain additional insight into these states, we perform a search for the $Z_{c}(4050)^{\pm}$ in $e^{+}e^{-}$ production using data collected by the BESIII 
experiment at center-of-mass energies between 4.18~GeV/$c^2$ and 4.60~GeV/$c^2$. 
The observation of other charged charmonium-like states, such as the $Z_{c}(3900)^{\pm}$ in $J/\psi\,\pi ^{+}\pi ^{-}$~\cite{PhysRevLett.110.252001} and  $Z_{c}(4020)^{\pm}$ in 
$h_{c}\pi ^{+}\pi ^{-}$~\cite{PhysRevLett.111.242001} in some of these data samples, make the BESIII experiment an ideal environment for the search for exotic particles.
In this paper the reaction channels $e^{+}e^{-} \rightarrow \chi _{cJ} \pi ^{+}\pi ^{-} $ ($J=0,1,2$) are investigated, in which the $Z_{c}(4050)^{\pm}$ resonance is expected to appear 
as a structure in the $\chi _{cJ}\pi ^{\pm} $ invariant-mass spectrum. Due to phase-space restrictions, the production of the second state $Z_{c}(4250)^{\pm}$ is only possible at higher energies. 

\section{Experimental Data and Monte Carlo samples}
The BESIII detector is a magnetic
spectrometer~\cite{BESIIISYS:2009} located at the Beijing Electron
Positron Collider (BEPCII)~\cite{Yu:IPAC2016-TUYA01}. The
cylindrical core of the BESIII detector consists of a helium-based
 multilayer drift chamber (MDC), a plastic scintillator time-of-flight
system (TOF), and a CsI(Tl) electromagnetic calorimeter (EMC),
which are all enclosed in a superconducting solenoidal magnet
providing a 1.0~T magnetic field. The solenoid is supported by an
octagonal flux-return yoke with resistive plate chamber muon-identifier 
modules interleaved with steel. The acceptance for
charged particles and photons is 93\% over the $4\pi$ solid angle. The
charged-particle momentum resolution at $1~{\rm GeV}/c$ is
$0.5\%$, and the $\mathrm{d}E/\mathrm{d}x$ resolution is $6\%$ for electrons
from Bhabha scattering. The EMC measures photon energies with a
resolution of $2.5\%$ ($5\%$) at $1$~GeV in the barrel (end-cap)
region. The time resolution of the TOF barrel part is 68~ps.
The time resolution of the end-cap TOF
system was upgraded in 2015 with multi-gap resistive plate chamber
technology, providing a time resolution of
60~ps. For data taken before 2015 the time resolution was 110~ps~\cite{etof::Guo2019,etof::Li2017}.

For the determination of reconstruction efficiencies and the estimation of background contributions, several Monte Carlo ({\sc MC}) simulated data samples were produced with a {\sc
geant4}-based~\cite{Agostinelli:2002hh} {\sc MC} software package. This
includes the geometric description of the BESIII detector and the
detector response. The signal channels  $e^{+}e^{-}\rightarrow \chi _{cJ} \pi ^{+}\pi ^{-} $, with $ \chi _{cJ} \rightarrow \gamma J/\psi $ ($J=0,1,2$) and $J/\psi\rightarrow \ell^{+}\ell^{-}$, 
are generated via the {\sc kkmc} generator~\cite{ref::kkmc} for the initial resonance and the event generator  {\sc evtgen}~\cite{ref::evtgen2001} for subsequent decays, using the phase-space 
distribution (PHSP). The PHSP model is also assumed for the decay $\chi _{cJ} \rightarrow \gamma J/\psi$, and the  VLL (vector to lepton lepton) model is used for the 
$J/\psi \rightarrow \ell^{+}\ell^{-}$ ($\ell = e,\mu$) decay. The generation of final state radiation is handled by the {\sc photos} package~\cite{photos}. 
The simulation includes the beam-energy spread and initial state radiation (ISR) in the $e^+e^-$
annihilations modeled with the generator {\sc kkmc}~\cite{ref::kkmc}. 
The inclusive MC samples consist of the production of open-charm
processes, the ISR production of vector charmonium(-like) states,
and the continuum processes incorporated in {\sc
kkmc}~\cite{ref::kkmc}.
Known decay modes are modeled with {\sc
evtgen} using branching fractions taken from the
Particle Data Group~\cite{pdg}, and the remaining unknown decays
from the charmonium states with {\sc
lundcharm}~\cite{ref:lundcharm1,ref:lundcharm2}. The size of these inclusive MC samples is scaled to five times of the integrated luminosity of their respective measured data point, with 
the exception at 4.18~GeV which has forty times the integrated luminosity.

The data sets studied in the analysis are shown in Table~\ref{datasets}.
Most of the samples correspond to an integrated luminosity~$\mathcal{L}$ of around 500 pb$^{-1}$. The samples taken at the center-of-mass energies of 4.18\,GeV, 4.23\,GeV, 4.26\,GeV and 4.42\,GeV are considerably larger.

Studies on {\sc MC} simulated samples are performed in order to optimize the event selection criteria.
For all generated {\sc MC} simulated signal samples of $e^{+}e^{-}\rightarrow \chi _{cJ} \pi^{+}\pi^{-}$, initial state radiation has been deactivated, except for 
those samples required for the dedicated investigation of the influence of ISR on the final result. Furthermore, several inclusive {\sc MC} samples have been analyzed to identify dominant background contributions. The dominant contributions found in the inclusive MC samples have been exclusively simulated. All generated exclusive {\sc MC} samples contain 500\,000 events.
\begingroup
\begin{table}[!ht]
\centering
\caption{\label{datasets}Data samples used in this analysis with the corresponding integrated luminosity $\mathcal{L}$~\cite{Lumi:2015}.}
\begin{tabular}{S[table-format=4.2]S[table-format=4.1]}
  \hline\hline
\multicolumn{1}{c}{$\sqrt{s}$~(MeV)} & \multicolumn{1}{c}{$\mathcal{L}$~(pb$^{-1}$)} \\ 
\hline
4178.00 & 3194.0 \\  
4189.27 & 526.7 \\  
4199.60 & 526.0 \\  
4209.72 & 517.1 \\  
4218.81 & 514.6 \\ 
4226.26 & 1056.4 \\ 
4235.83 & 530.3 \\ 
4243.89 & 538.1 \\ 
4257.97 & 828.4 \\ 
4266.93 & 531.1 \\ 
4277.79 & 175.7 \\
4358.26 & 543.9 \\  
4415.58 & 1044.0 \\  
4527.14 & 112.1 \\
4599.53 & 586.9 \\ 
\hline
\end{tabular}
\end{table}
\endgroup
\FloatBarrier

\section{\label{sec::PID}Event selection}
Several selection criteria are applied in order to perform the particle identification (PID) and the event selection.

Photon candidates are constructed from clusters of energy deposits of at least 25\,MeV of energy in the barrel part of the EMC (polar angle region of $|\cos \theta| < 0.80$ with respect to the beam axis) and 50\,MeV in the endcap region ($0.86 <|\cos \theta| < 0.92$).
The corresponding EMC time is required to be within a window of 700\,ns relative to the event start time, and the candidates 
are requested to be at least  20$^{\circ}$ away from the nearest charged track to reject EMC hits caused by split-offs of clusters of charged particles.

Charged-track candidates must pass the interaction point within a cylindrical volume, with a radius of 1\,cm and length of 10\,cm, around the interaction point. 
Furthermore, due to the limitation of the MDC acceptance, the region close to the beams is excluded by 
requiring $|\cos \theta_{\text{Track}}| < 0.93$.
To distinguish pion candidates from the leptons coming from the $J/\psi$, a combination of the track momenta measured with the MDC ($P_{\text{MDC}}$) and the energy deposited in the EMC ($E_{\text{EMC}}$) is used.
Pion candidates are tracks  with a momentum smaller than 1.0\,GeV/$c$ and lepton candidates are tracks with a momentum greater than 1.0\,GeV/$c$.
Furthermore, to separate the electrons from muons, tracks with a ratio of $E_{\text{EMC}}/P_{\text{MDC}}<0.3$\,$c$ are considered to be muon candidates and tracks with $E_{\text{EMC}}/P_{\text{MDC}}>0.7$\,$c$ 
are considered as electron candidates.

In order to select $e^{+}e^{-}\rightarrow \chi _{cJ} \pi^{+}\pi^{-}$ events, four track candidates with a net charge of zero, including two lepton candidates, and at least one photon candidate are required. 
A vertex fit of the tracks to a common vertex is applied. 
Then, a kinematic fit with constraints on the initial-four-momentum (4C) and the mass of the  $J/\psi$ meson (5C-fit) to $m_{J/\psi,\text{PDG}}$~\cite{pdg} is performed.
Candidates that satisfy $\chi ^{2}_{\text{5C}} < 50$ are retained for further analysis.
If multiple candidates are found in an event, the one with the lowest $\chi ^{2}_{\text{5C}}$ value is selected. 
However, only one candidate is seen after the event selection in signal MC data and predominantly one in data. 

\section{\label{sec::BKGDATA}Background studies}

The following processes have been identified as the principal sources of background events through the study of the inclusive Monte Carlo samples:

\noindent
	$e^{+}e^{-} \rightarrow e^{+}e^{-}\gamma_{\text{ISR}},\,\gamma_{\text{ISR}} \rightarrow e^{+}e^{-}$;\\
	$e^{+}e^{-} \rightarrow \eta J/\psi,\,\eta\rightarrow \gamma\,\pi^{+}\pi^{-}$;\\
	$e^{+}e^{-} \rightarrow \eta' J/\psi,\,\eta'\rightarrow \gamma\,\rho^{0},\rho^{0}\rightarrow \pi^{+}\pi^{-}$;\\
	$e^{+}e^{-} \rightarrow \omega\chi_{cJ},\,\omega\rightarrow\pi^{+}\pi^{-},\,\chi_{cJ}\rightarrow\gamma  J/\psi$ ($J=0,1,2$);\\
	$e^{+}e^{-} \rightarrow \gamma_{\text{ISR}}\,\psi(2S),\,\psi(2S)\rightarrow \pi^{+}\pi^{-} J/\psi$; and\\
	$e^{+}e^{-} \rightarrow Y(4260) \rightarrow \gamma \,X(3872),\,X(3872)\rightarrow J/\psi\,\pi^{+}\pi^{-}$.\\	
In all these reactions the $J/\psi$ decays into a lepton pair ($e^{+}e^{-}/\mu^{+}\mu^{-}$). Apart from the first process, these contributions have the same 
final state as the signal reaction channel and are thus not distinguishable by the applied kinematic fit.
Additional selection criteria based on other kinematic variables are required to suppress these background channels.\\
\begin{figure}[!htbp]
\includegraphics[width=8.15cm]{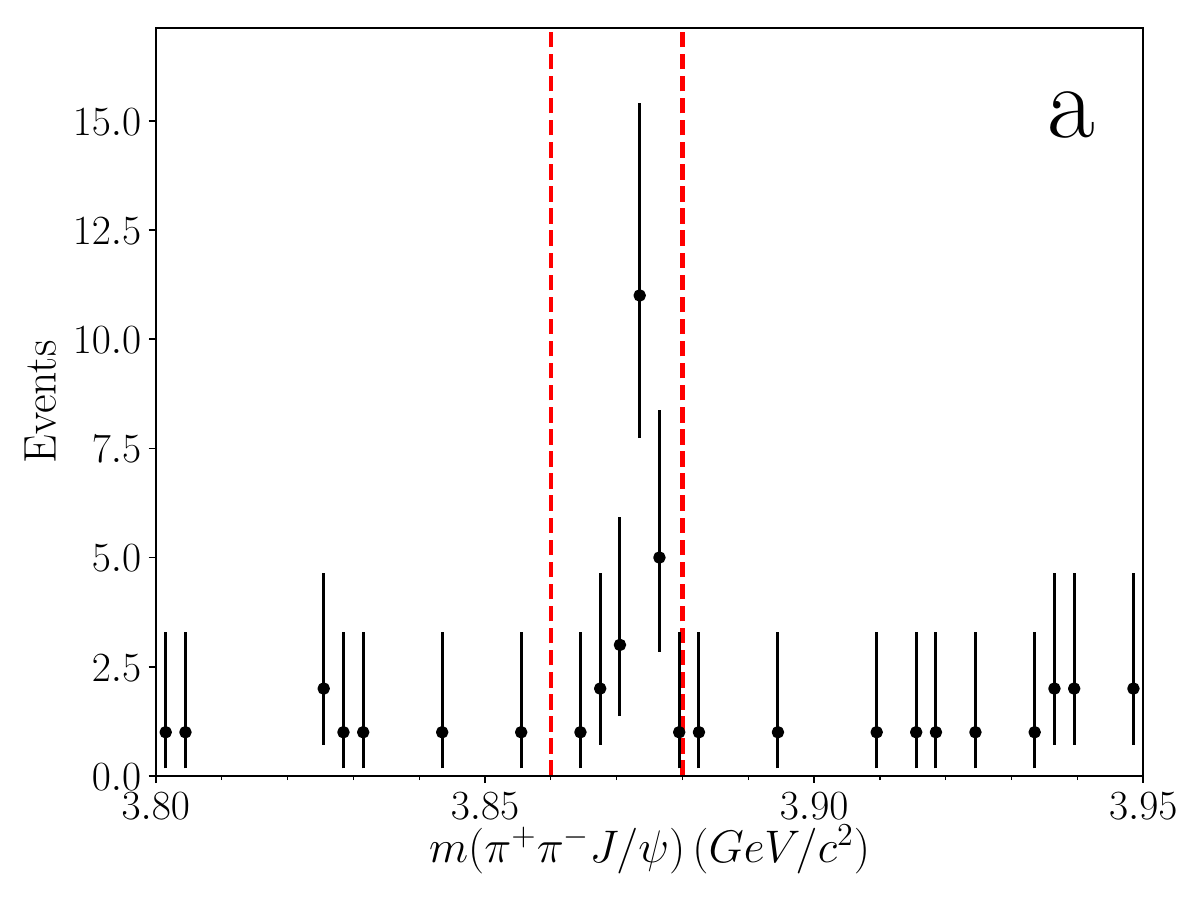}
\includegraphics[width=8.15cm]{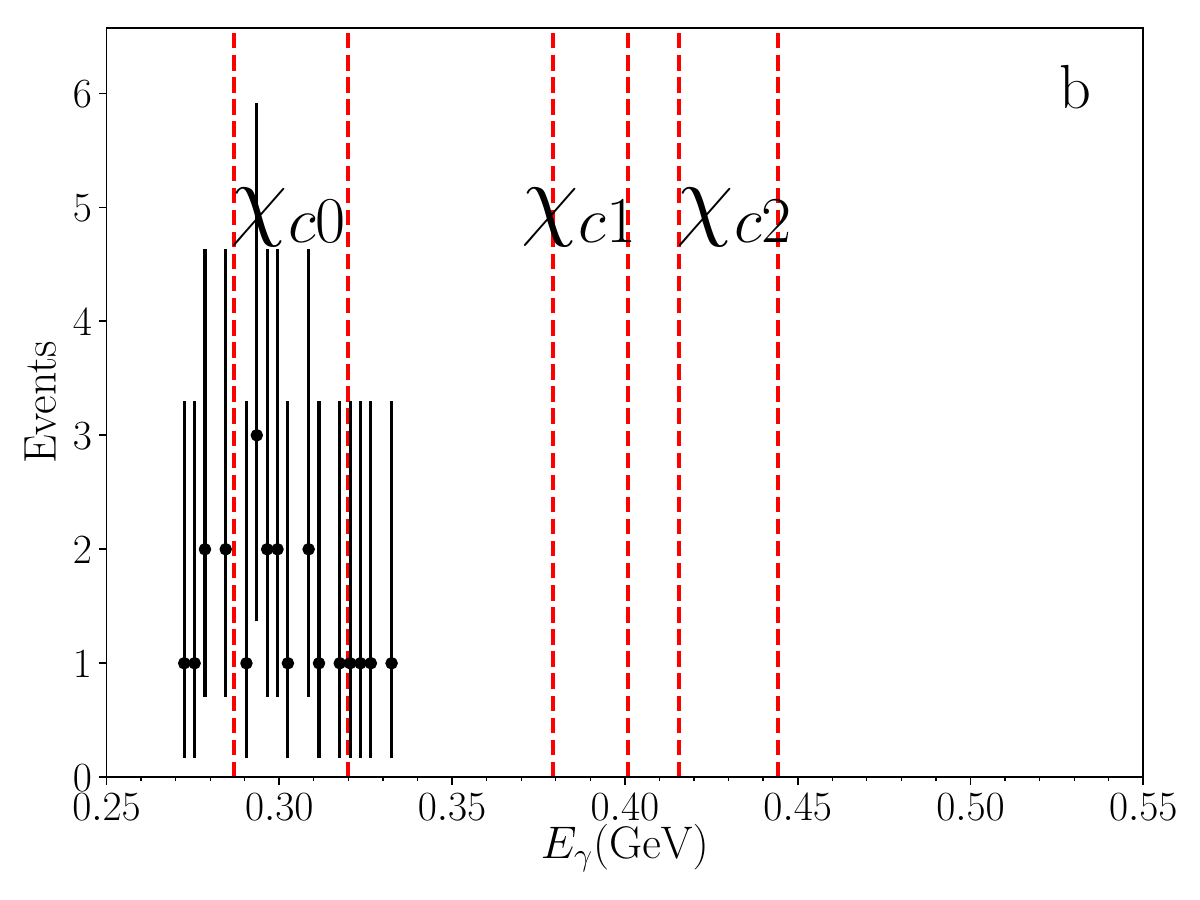}
\caption{Contamination from $e^{+}e^{-} \rightarrow Y(4260) \rightarrow \gamma \,X(3872),\,X(3872)\rightarrow \pi^{+}\pi^{-} J/\psi$ events at 4.18\,GeV. Plot (a) shows the $X(3872)$ signal, with a selection window indicated by red dashed lines to isolate the events in plot (b), which shows the photon energy.  Here, the red dashed lines indicate the $\chi_{cJ}$ selection windows}
\label{figure::X3872}
\end{figure}
Background from Bhabha scattering with associated ISR/FSR photons that convert into an electron-positron pair misidentified as a pion pair is suppressed by the requirement that the pion opening angle in the laboratory system, $\alpha ^{\pi ^{+}}_{\pi ^{-}}$, satisfies  $\cos (\alpha ^{\pi ^{+}}_{\pi ^{-}}) < 0.98 $. This criterion results in a signal loss below 1\% for all studied energy points.
The background contributions from $\eta J/\psi$  and $\eta ' J/\psi$ are rejected
by requiring $m_{\text{rec}} \geq 0.57$\,GeV/$c^{2}$ and rejecting candidates with $0.95\leq m_{\text{rec}}  \leq 0.97 $\,GeV/$c^{2}$, where $m_{\text{rec}}$ is
the $J/\psi$ recoil mass.
Contamination from $\omega \chi _{cJ}$ events are suppressed by
rejecting candidates where the $\chi _{cJ}$ recoil mass lies between $0.74$ and $0.82$\,GeV/$c^{2}$.

The main ISR background contribution originates from the process
$e^{+}e^{-} \rightarrow \gamma _{\text{ISR}} \psi(2S)$.  This reaction is dangerous because the ISR photon has a wide range of possible energies, depending on the center of mass energy.
The final source of contamination that is considered  is illustrated in Fig.~\ref{figure::X3872}, where events that are 
most likely coming from $\gamma X(3872)$ are confused with $\pi^{+}\pi^{-}\chi_{c0}$ 
signal candidates. Apparent $Y(4260) \rightarrow \gamma \,X(3872)$ events at the center-of-mass energy of 4.18\,GeV 
coincidentally have a photon energy similar to the one coming from a radiative decay of $\chi_{c0}\rightarrow \gamma\,J/\psi$.

Exclusive Monte Carlo data sets containing 500\,000 events each are simulated and analyzed for each background process and center-of-mass energy.
For $e^{+}e^{-} \rightarrow \gamma _\text{ISR} \psi(2S)$, samples are simulated for each studied center-of-mass energy 
using {\sc kkmc} to evaluate the cross section.
Events coming from $e^{+}e^{-} \rightarrow \gamma _{\text{ISR}} \psi (2S)/\gamma X(3872)$ with
$\psi (2S)/X(3872) \rightarrow \pi ^{+}\pi ^{-} J/\psi$ decays are suppressed by rejecting the region
$m_{\pi ^{+}\pi ^{-} J/\psi} \leq 3.71$\,GeV/$c^{2}$ 
and $3.86$\,GeV/$c^{2}$ $\leq  m_{\pi ^{+}\pi ^{-} J/\psi} \leq 3.88$\,GeV/$c^{2}$, respectively.
Due to the restrictions on the phase space, $\omega \chi _{c1}$ and $\omega \chi _{c2}$ contribute only for center-of-mass energies above 4.3\,GeV and $\omega \chi _{c0}$ only above 4.2\,GeV, respectively. 

The reconstruction efficiency is 
evaluated with simulated MC data of the signal channel.
The reconstruction efficiency ranges from 16 to 28\% after the application of all selection criteria, depending on the center-of-mass energy (see \cref{table::results0,table::results1,table::results2}).

\section{\label{sec::UL}Cross-Section Determination}

\begin{figure}[!t]
\includegraphics[width=8.6cm]{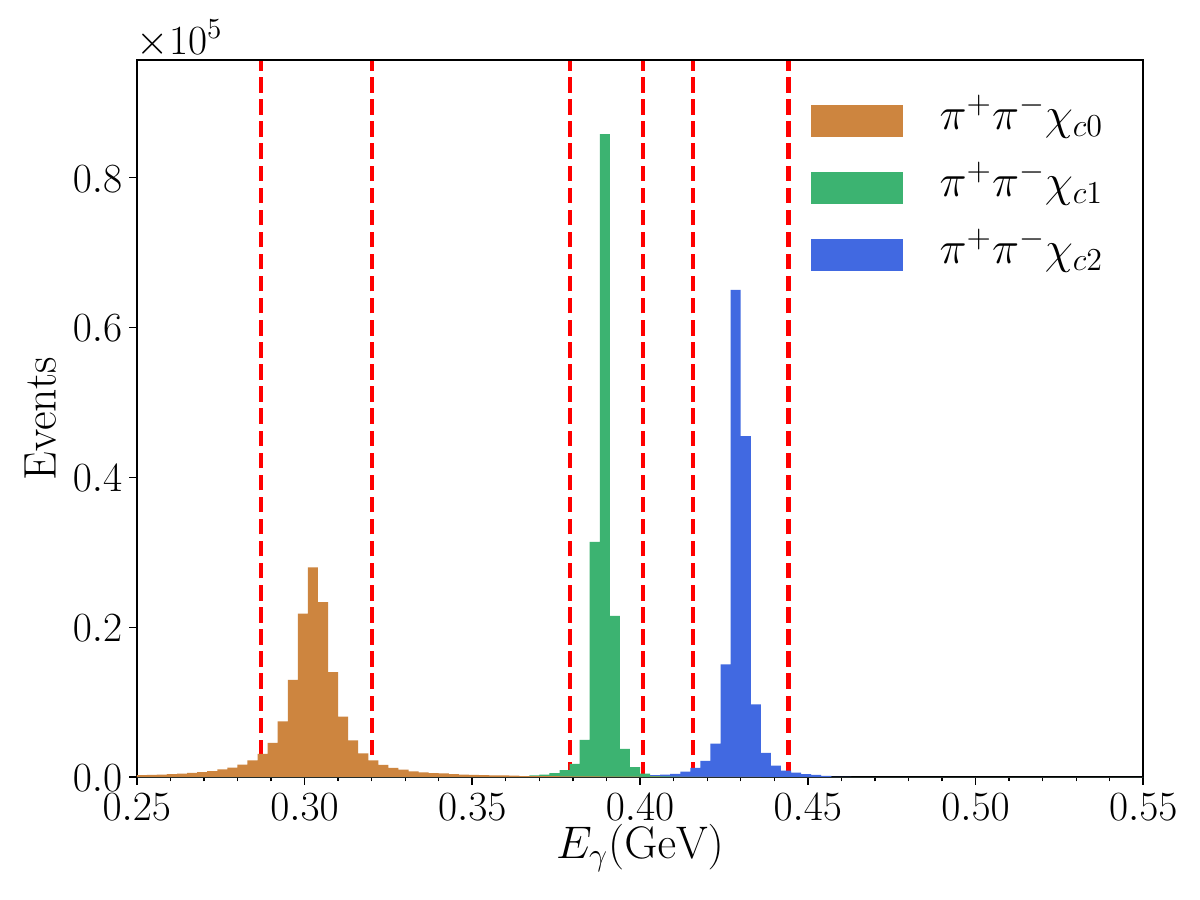}
\caption{\label{figure::photon_energy} Reconstructed photon energy $E_{\gamma}$ of the $\chi _{cJ}$ candidates measured in the rest frame of the $\pi^{+}\pi^{-}$ recoil system 
	from generated $\chi _{cJ}\pi^{+}\pi^{-}$ Monte Carlo data sets. The red dashed lines indicate the selection windows. The histograms are normalized to the same integral.}
\end{figure}

The signal yield is directly determined by counting the events surviving the selection criteria.
Since the radiative process $\chi _{cJ}\rightarrow \gamma J/\psi$ is a two-body decay, the photon energy of each decay mode 
serves as a distinctive signature for the separation of
the three $\chi_{cJ}$ channels. Figure~\ref{figure::photon_energy} shows 
the photon energy after boosting it into the $\pi^{+}\pi^{-}$ recoil system. 
This method allows for a clear separation of the three $\chi_{cJ}$ channels by setting the 
(boosted) photon energy windows and leads to the results shown in \cref{table::results0,table::results1,table::results2}. There, the first uncertainties are 
statistical and the second systematic, arising from the sources discussed in Section~\ref{sec::SYSTEMATICS}. 
The expected background events for each center-of-mass energy are estimated 
by adding up each background contribution:
\begin{equation}
N_{\text{bkg}}~\equiv~\mathcal{L}~\sum_{i}~\sigma_{i}~\mathcal{B}_{i}~\epsilon_{i}~,  
\end{equation}
where $\mathcal{L}$ is the integrated luminosity at a given center-of-mass energy, $\sigma_{i}$ is the cross section for each background contribution, 
$\mathcal{B}_{i}$ the corresponding branching ratio and $\epsilon_{i}$ the efficiency from the exclusive background {\sc MC} data samples after all 
selection criteria. The values of $\sigma_{i}$ are taken from previous BESIII  
measurements~\cite{PhysRevD.94.032009,PhysRevD.91.112005,PhysRevLett.114.092003,PhysRevD.93.011102,PhysRevD.99.091103}.  In the cases where no cross section has yet been measured the upper limits are used to provide an estimate. 
Finally, $\mathcal{B}_{i}$ is taken from the Particle Data Group (PDG)~\cite{pdg}. 

The observed cross-section $\sigma_{\text{obs}}$ is calculated via
\begin{equation}
 \sigma_{\text{obs}}~\equiv~\frac{N_{\text{obs}} - N_{\text{bkg}}}{\mathcal{L}~\epsilon~\mathcal{B}({\chi_{cJ} \rightarrow \gamma J/\psi})~\mathcal{B}({J/\psi \rightarrow \ell^{+} \ell^{-}})}~,
 \label{equ::sigma}
\end{equation}
with the selection efficiency
 $\epsilon$ and $\mathcal{B}({\chi_{cJ} \rightarrow \gamma J/\psi})$ being the corresponding branching fraction for the selected $\chi_{cJ}$ 
decay channel and $\mathcal{B}({J/\psi \rightarrow \ell^{+} \ell^{-}})$ the sum of the two branching fractions $\mathcal{B}({J/\psi \rightarrow e^{+} e^{-}})$ and $\mathcal{B}({J/\psi \rightarrow \mu^{+} \mu^{-}})$.

The determination of the upper limits is discussed in further detail in Section~\ref{sec::SYS_INTERPRETATION}. 

\section{\label{sec::SYSTEMATICS} Systematic-uncertainty estimation}

Systematic uncertainties are assigned, where appropriate, for each step and input in the analysis.
The uncertainty on the measurement of the integrated luminosity is
1\%~\cite{Lumi:2015}. The uncertainty on the reconstruction efficiency due to the finite size of the MC simulation sample is 0.3-0.4\%. The difference between data and MC simulation of the track and photon reconstruction efficiencies and also the correlation between the tracks are taken into account by assigning a 1\% uncertainty per track~\cite{tracks::2014} and per photon~\cite{photon::2011},  resulting in an overall uncertainty of 4.1\%. The uncertainty associated with final state radiation is stated to be roughly 0.1\%~\cite{BARBERIO1994291} and considered to be negligible.

The uncertainty associated with the selection criteria is assigned to be the largest shift in efficiency observed when the applied criteria are moved by 10\% in both directions.
For the selection on the $\chi^{2}_{\text{5C}}$ of the kinematic fit, this results in an uncertainty of around 1.4\%, depending on the center-of-mass energy and applied $\chi_{cJ}$ selection.
For the $\eta$ veto the range is much larger and varies between 0.2\% and 4.5\%.
 
\begin{table}[!b]
\centering
\caption{\label{table::totSys}Systematic uncertainties separated for the different reaction channels $e^{+}e^{-}\rightarrow\chi_{cJ}\pi^{+}\pi^{-}$. Contributions vary depending on the center-of-mass energy.}
\resizebox{8.6cm}{!}{
\begin{tabular}{lccc}
\hline
\hline 
 Source & $\sigma_{\text{sys}}$,$\chi_{c0}$(\%)  & $\sigma_{\text{sys}}$,$\chi_{c1}$(\%)  & $\sigma_{\text{sys}}$,$\chi_{c2}$(\%)\\
\hline 
Luminosity &   1.0 &   1.0 &   1.0\\  
Rec. eff.&  0.3 - 0.4  &  0.3 - 0.4  &  0.3 - 0.4 \\ 
Track/photon &   4.1 &   4.1 &   4.1\\  
\hline 
$\chi ^{2}$-veto &  1.3 - 1.9 & 1.3 - 1.7   &  1.3 - 1.8  \\ 
$\eta$-veto & 0.6 - 3.7   & 0.4 - 2.7   &  0.2 - 4.5  \\ 
\hline 
$\pi ^{+}\pi ^{-}$-angle &  0.4 - 0.5  &  0.4 - 0.5  &  0.3 - 0.5 \\ 
$\psi (2S)$-veto &  0.0   & 0.0 - 2.4   &  0.0 - 9.5 \\ 
$\eta '$-veto & 0.2 - 1.1   & 0.2 - 1.2   &   0.3 - 1.3 \\ 
$\omega$-veto &  0.3 - 1.9  &  0.0 - 1.8  &   0.0 - 1.8\\ 
X(3872)-veto&  0.0 - 4.1  &  0.0 - 2.8  &  0.0 - 2.4 \\ 
\hline 
$\chi _{cJ}$-selection & 4.7 - 5.2  &   0.0 - 0.0 &   0.0 - 0.1\\  
\hline 
Total &  6.9 - 8.7  &  4.7 - 5.9  &  4.7 - 11.0 \\ 
\hline
\hline
\end{tabular}
}
\end{table}

Similarly, uncertainties associated with other selection criteria also depend on the collision energy.
For the background vetoes, the windows are increased and decreased by 10\% and again the largest difference, which varies in the range of a few percent, is assigned. In the case of the $\chi_{c2}$ selection, the $\psi(2S)$ veto contributes larger systematic uncertainties at lower center-of-mass energies, where the invariant 
$\pi^{+}\pi^{-}J/\psi$ mass of the expected signal lies, coincidentally, in the vicinity of the $\psi(2S)$ mass. 
The systematic uncertainty is largest for the $\chi_{c0}$ selection, on account of the larger natural width of this state.

Table~\ref{table::totSys} summarizes the individual systematic uncertainties.
Contributions arising from the variation of a certain input from the nominal value are considered to be negligible if the observed change in result is found to be less than the uncorrelated systematic uncertainty.  The total systematic uncertainty is calculated as the sum in quadrature of each component, assuming negligible correlations, and results in values between 4.7\% to 11.0\%.
When calculating upper limits, a Gaussian-shaped uncertainty is added to the efficiency with a width equal to the total systematic uncertainty.

\begin{table*}[!ht]
\centering
\caption{\label{table::results0} Measured cross sections and associated information for $e^{+}e^{-}\rightarrow\chi _{c0}\pi^{+}\pi^{-}$ at different center-of mass-energies $E_{\text{cms}}$. Shown are the integrated luminosity $\mathcal{L}$, the selection efficiency
 $\epsilon$, the number of observed events $N_{\text{obs}}$, the number of expected background events $N_{\text{bkg}}$, 
 the observed cross sections $\sigma _{\text{obs}}$ with statistical and systematic uncertainties, the statistical significance and the respective upper limits at 90\% confidence level. }

\newcommand\ST{\rule[-0.2em]{0pt}{1.2em}}

\begin{ruledtabular}
\begin{tabular}{cccccccc}
  \multicolumn{1}{c}{$E_{\text{cms}}$\,(GeV)}
  & \multicolumn{1}{c}{$\mathcal{L}$\,($\text{pb}^{-1}$)}
  & $\epsilon$\,(\%)
  & $N_{\text{obs}}$
  & $N_{\text{bkg}}$ 
  & $\sigma _{\text{obs}}$\,(pb)
  & significance\,($\sigma$)
  & $\sigma _{\text{UL}}$\,(pb)\\ 
\hline 
4.178 & 3194.0 & 16.21 & 3 & 0.0 & 3.47 $_{-2.26}^{+3.59}$ $\pm$ 0.30 & 1.15  & 11.8 \ST \\
4.189 & 526.7 & 16.43 & 1 & 0.0 & 6.92 $_{-6.23}^{+16.10}$ $\pm$ 0.60 & 0 &  37.7 \ST \\
4.200 & 526.0 & 16.31 & 0 & 0.0 & 0 $_{-0}^{+8.01}$ $\pm$ 0 & 0 & 20.6 \ST \\
4.210 & 517.1 & 16.38 & 1 & 0.0 & 7.07 $_{-6.37}^{+16.40}$ $\pm$ 0.58 & 0 & 38.4  \ST\\
4.219 & 514.6 & 16.72 & 0 & 0.0 & 0 $_{-0}^{+7.99}$ $\pm$ 0 & 0  & 20.5  \ST\\
4.226 & 1056.0 & 17.01 & 3 & 0.0 & 9.99 $_{-6.50}^{+10.40}$ $\pm$ 0.80 & 1.15  & 34.0  \ST\\
4.236 & 530.3 & 18.14 & 0 & 0.0 & 0 $_{-0}^{+7.14}$ $\pm$ 0 & 0 & 18.4  \ST\\
4.244 & 538.1 & 19.02 & 3 & 0.0 & 17.60 $_{-11.40}^{+18.20}$ $\pm$ 1.32 & 1.15 & 59.6 \ST \\
4.258 & 828.4 & 19.70 & 2 & 0.0 & 7.34 $_{-5.41}^{+10.00}$ $\pm$ 0.55 & 0.67 & 29.1 \ST \\
4.267 & 531.1 & 21.10 & 2 & 0.0 & 10.70 $_{-7.88}^{+14.60}$ $\pm$ 0.77 & 0.67 & 42.4 \ST \\
4.278 & 175.7 & 21.29 & 0 & 0.0 & 0 $_{-0}^{+18.40}$ $\pm$ 0 & 0 & 47.3  \ST\\
4.358 & 543.9 & 21.58 & 1 & 0.0 & 5.10 $_{-4.60}^{+11.90}$ $\pm$ 0.36 & 0 & 27.8  \ST\\
4.416 & 1044.0 & 21.86 & 0 & 0.0 & 0 $_{-0}^{+3.01}$ $\pm$ 0 & 0 & 7.8  \ST\\
4.527 & 112.1 & 23.85 & 0 & 0.0 & 0 $_{-0}^{+25.70}$ $\pm$ 0 & 0 & 66.1  \ST\\
4.600 & 586.9 & 23.92 & 2 & 0.0 & 8.50 $_{-6.29}^{+11.70}$ $\pm$ 0.61 & 0.67 & 33.8 \ST \\
\end{tabular}
\end{ruledtabular}

\caption{\label{table::results1} Measured cross-sections and associated information for $e^{+}e^{-}\rightarrow\chi _{c1}\pi^{+}\pi^{-}$. See table\,~\ref{table::results0} for more information.}

\begin{ruledtabular}
\begin{tabular}{cccccccccc}
  \multicolumn{1}{c}{$E_{\text{cms}}$\,(GeV)}
  & \multicolumn{1}{c}{$\mathcal{L}$\,($\text{pb}^{-1}$)}
  & $\epsilon$\,(\%)
  & $N_{\text{obs}}$
  & $N_{\text{bkg}}$ 
  & $\sigma _{\text{obs}}$\,(pb)
  & significance\,($\sigma$)
  & $\sigma _{\text{UL}}$\,(pb)\\ 
\hline 
4.178 & 3194.0 & 26.36 & 2 & 0.0 & 0.06 $_{-0.04}^{+0.08}$ $\pm$ 0 & 0.67  & 0.23  \ST\\
4.189 & 526.7 & 27.16 & 0 & 0.0 & 0 $_{-0}^{+0.20}$ $\pm$ 0 & 0 & 0.50 \ST \\
4.200 & 526.0 & 27.28 & 0 & 0.0 & 0 $_{-0}^{+0.20}$ $\pm$ 0 & 0 & 0.50  \ST\\
4.210 & 517.1 & 27.24 & 1 & 0.12 & 0.17 $_{-0.14}^{+0.40}$ $\pm$ 0 & 0 & 0.94 \ST \\
4.219 & 514.6 & 27.24 & 0 & 0.0 & 0 $_{-0}^{+0.2}$ $\pm$ 0 & 0 & 0.51  \ST\\
4.226 & 1056.4 & 26.03 & 4 & 0.0 & 0.36 $_{-0.17}^{+0.28}$ $\pm$ 0.02 & 1.53 & 1.09  \ST\\
4.236 & 530.3 & 24.71 & 2 & 0.0 & 0.37 $_{-0.24}^{+0.49}$ $\pm$ 0.02 & 0.67 & 1.47  \ST\\
4.244 & 538.1 & 23.36 & 2 & 0.0 & 0.39 $_{-0.25}^{+0.51}$ $\pm$ 0.02 & 0.67 & 1.53  \ST\\
4.258 & 828.4 & 21.56 & 2 & 0.0 & 0.27 $_{-0.18}^{+0.36}$ $\pm$ 0.02 & 0.67  & 1.08  \ST\\
4.267 & 531.1 & 22.32 & 0 & 0.0 & 0 $_{-0}^{+0.24}$ $\pm$ 0 & 0 & 0.61  \ST\\
4.278 & 175.7 & 22.19 & 0 & 0.0 & 0 $_{-0}^{+0.72}$ $\pm$ 0 & 0 & 1.85  \ST\\
4.358 & 543.9 & 23.48 & 1 & 0.0 & 0.19 $_{-0.16}^{+0.44}$ $\pm$ 0 & 0 & 1.04 \ST \\
4.416 & 1044.0 & 25.19 & 0 & 0.0 & 0 $_{-0}^{+0.11}$ $\pm$ 0 & 0 & 0.28 \ST \\
4.527 & 112.1 & 27.61 & 0 & 0.0 & 0 $_{-0}^{+0.91}$ $\pm$ 0 & 0 & 2.33  \ST\\
4.600 & 586.9 & 27.72 & 2 & 0.0 & 0.3 $_{-0.19}^{+0.40}$ $\pm$ 0.02 & 0.67 & 1.18  \ST\\
\end{tabular}
\end{ruledtabular}

\caption{\label{table::results2} Measured cross-sections and associated information for $e^{+}e^{-}\rightarrow\chi _{c2}\pi^{+}\pi^{-}$. See table\,~\ref{table::results0} for more information.}

\begin{ruledtabular}
\begin{tabular}{cccccccccc} 
  \multicolumn{1}{c}{$E_{\text{cms}}$\,(GeV)}
  & \multicolumn{1}{c}{$\mathcal{L}$\,($\text{pb}^{-1}$)}
  & $\epsilon$\,(\%)
  & $N_{\text{obs}}$
  & $N_{\text{bkg}}$ 
  & $\sigma _{\text{obs}}$\,(pb)
  & significance\,($\sigma$)
  & $\sigma _{\text{UL}}$\,(pb)\\ 
\hline 
4.178 & 3194.0 & 16.90 & 4 & 2.02 & 0.16 $_{-0.16}^{+0.26}$ $\pm$ 0.02 & 0.40 & 0.82  \ST\\
4.189 & 526.7 & 19.26 & 1 & 0.0 & 0.44 $_{-0.36}^{+1.00}$ $\pm$ 0.04 & 0 & 2.38 \ST \\
4.200 & 526.0 & 21.51 & 0 & 0.0 & 0 $_{-0}^{+0.45}$ $\pm$ 0 & 0 & 1.15 \ST \\
4.210 & 517.1 & 23.59 & 0 & 0.0 & 0 $_{-0}^{+0.42}$ $\pm$ 0 & 0 & 1.07  \ST\\
4.219 & 514.6 & 25.21 & 1 & 0.0 & 0.34 $_{-0.28}^{+0.78}$ $\pm$ 0.02 & 0 & 1.84 \ST \\
4.226 & 1056.0 & 25.61 & 3 & 0.0 & 0.49 $_{-0.27}^{+0.48}$ $\pm$ 0.03 & 1.15 & 1.65  \ST\\
4.236 & 530.3 & 27.29 & 0 & 0.0 & 0 $_{-0}^{+0.35}$ $\pm$ 0 & 0 & 0.90  \ST\\
4.244 & 538.1 & 27.90 & 1 & 0.0 & 0.29 $_{-0.24}^{+0.68}$ $\pm$ 0.01 & 0 & 1.59 \ST \\
4.258 & 828.4 & 26.59 & 1 & 0.0 & 0.2 $_{-0.17}^{+0.46}$ $\pm$ 0.01 & 0 & 1.09  \ST\\
4.267 & 531.1 & 27.00 & 0 & 0.0 & 0 $_{-0}^{+0.35}$ $\pm$ 0 & 0 & 0.91  \ST\\
4.278 & 175.7 & 25.19 & 1 & 0.0 & 1.0 $_{-0.83}^{+2.29}$ $\pm$ 0.05 & 0 & 5.4  \ST\\
4.358 & 543.9 & 21.54 & 0 & 0.0 & 0 $_{-0}^{+0.43}$ $\pm$ 0 & 0 & 1.12  \ST\\
4.416 & 1044.0 & 23.91 & 2 & 0.0 & 0.35 $_{-0.23}^{+0.47}$ $\pm$ 0.02 & 0.67 & 1.4 \ST \\
4.527 & 112.1 & 27.23 & 0 & 0.0 & 0 $_{-0}^{+1.66}$ $\pm$ 0 & 0 & 4.26  \ST\\
4.600 & 586.9 & 27.27 & 0 & 0.0 & 0 $_{-0}^{+0.32}$ $\pm$ 0 & 0 & 0.81 \ST \\
\end{tabular}
\end{ruledtabular}

\end{table*}

\section{\label{sec::ISRCORRECTION}ISR correction}

An ISR correction factor is applied to the measured cross section, as listed in~\cref{table::results0,table::results1,table::results2}.
 The number of observed events can be written as
\begin{equation}
N = \mathcal{L} \int \sigma (x) \epsilon (x) W(x) dx
\end{equation}
where $x \equiv E_{\text{ISR}}/E_{\text{beam}}$ and $W(x)$ is the radiator function~\cite{MONTAGNA199931,BESIIISYS:2014}.
After factoring out the Born cross section $\sigma _{0}$ and the efficiency $\epsilon _{0}$ at $x=0$ this expression becomes
\begin{equation}
N = \mathcal{L} \sigma _{0} \epsilon _{0} \int \frac{\sigma (x)}{\sigma _{0}} \frac{\epsilon (x)}{\epsilon _{0}} W(x) dx~.
\end{equation}
The ISR correction factor is defined as
\begin{equation}
\kappa \equiv \int \frac{\sigma (x)}{\sigma _{0}} \frac{\epsilon (x)}{\epsilon _{0}} W(x) dx 
\end{equation}
so that
\begin{equation}
N = \mathcal{L} \sigma _{0} \epsilon _{0} \kappa .
\end{equation}

\begin{figure}[!htbp]
\includegraphics[width=8.15cm]{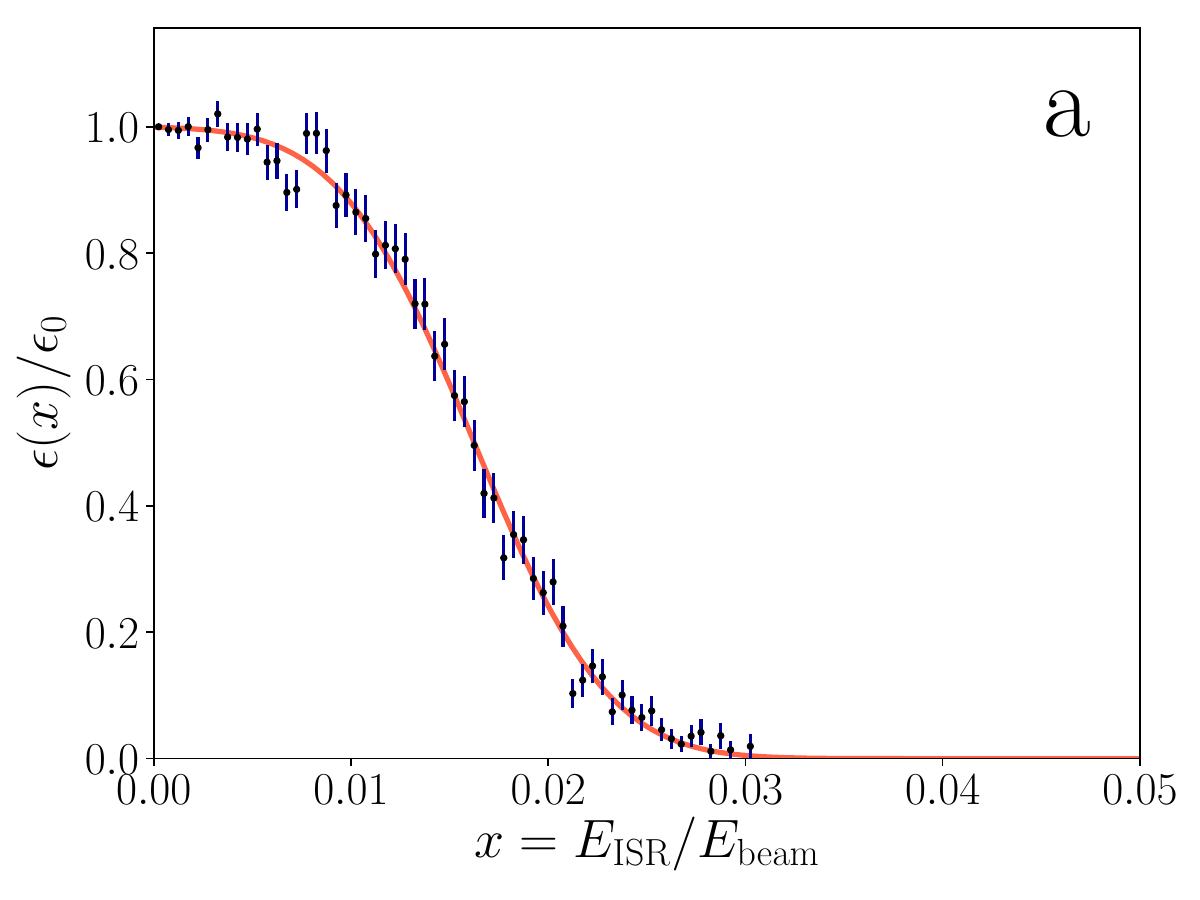}
\includegraphics[width=8.15cm]{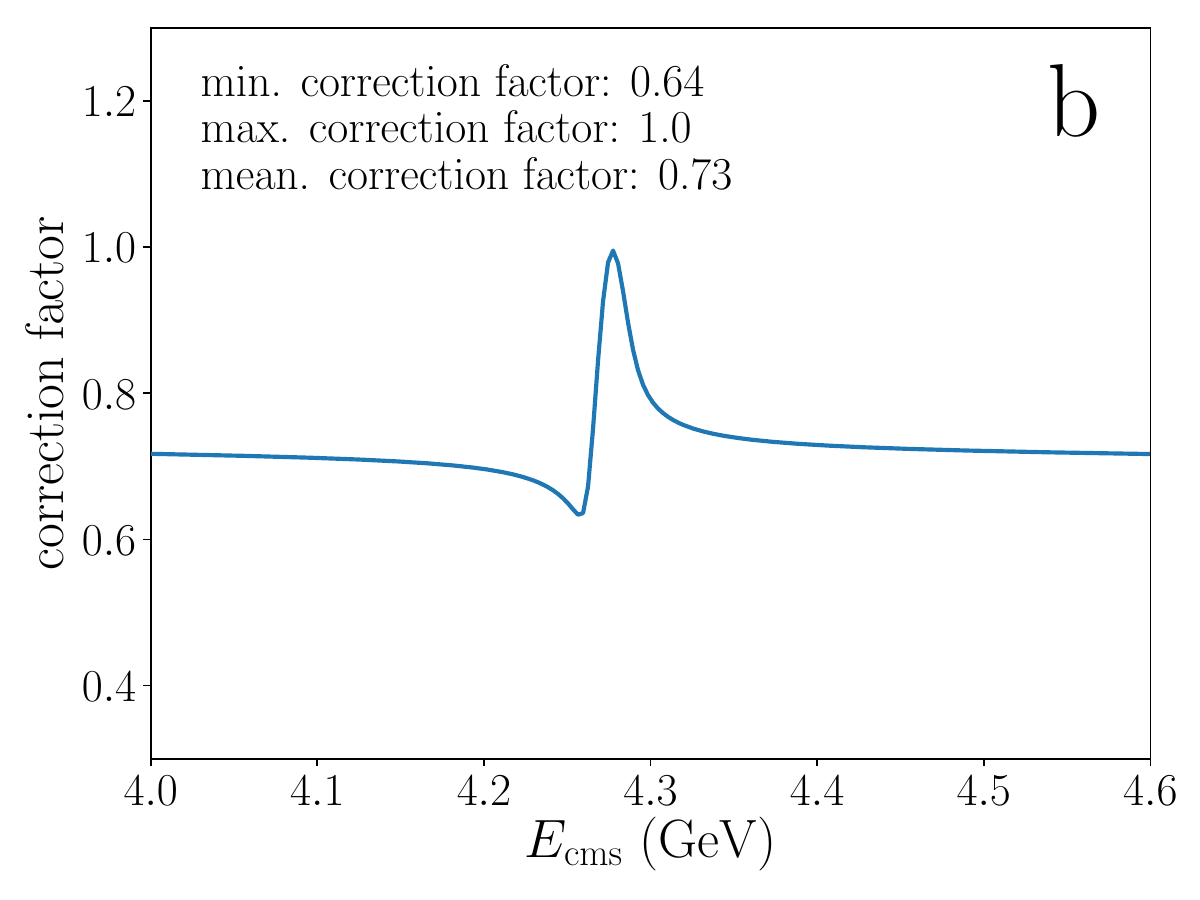}
\caption{\label{figure::ISRFactorDependency}ISR correction for the reaction channel $e^{+}e^{-}\rightarrow\pi^{+}\pi^{-}\chi _{c1}$ at $E_{\text{cms}}=4.6$\,GeV. ($a$) shows the normalized reconstruction efficiency versus the normalized energy of the ISR photon $E_{\text{ISR}}/E_{\text{beam}}$, where the red curve represents a fit by the error function;
($b$) shows the dependence of the ISR correction factor on $E_{\text{cms}}$, assuming a single narrow resonance with mass of  4.26\,GeV/$c^2$ and width of 10\,MeV.}
\end{figure}

The efficiency ratio $\epsilon (x)/ \epsilon _{0}$ is determined from a sample of  MC simulated signal events, which are generated including  ISR. 
Figure~\ref{figure::ISRFactorDependency} $a$ shows the efficiency ratio as a function of $x$ for the $\chi _{c1}$ signal MC sample at 4.6\,GeV.  The superimposed fit is an error function, 
which is found to describe all  $\chi _{cJ}$ modes and collision energies.   

The correction factor $\kappa$  is strongly correlated to the energy dependence
of the signal cross section, which is currently unknown. To obtain
conservative upper limits on the signal we estimate the lowest possible
$\kappa$ value.  We assume a narrow resonance of width 10\,MeV and mass  4.26\,GeV/$c^2$, which results in the $\kappa$
energy dependence shown in Fig.~\ref{figure::ISRFactorDependency} $b$. 
Changing the position of the
resonance results in a corresponding shift of the $\kappa$ energy
dependence, while the shape is nearly unchanged. The minimal value of   
the correction factor, $\kappa = 0.64$, is conservatively used to set the
upper limits of the cross section at all collision energies.

\section{\label{sec::SYS_INTERPRETATION} Upper-limit determination}
The upper limits on the branching ratios are calculated following a frequentist procedure~\cite{COUSINS2008480,Rolke:2009},
using the definition
\begin{equation}
 \sigma_{\text{UL}}~\equiv~\frac{N_{\text{UL}}}{\mathcal{L}\;(1+\delta)\;\frac{1}{|1-\Pi(s)|^{2}}~\epsilon~\mathcal{B}}~.
 \label{equ::sigma_UL}
\end{equation}
Here  $N_{\text{UL}}$ is  the upper limit on the signal yield,
$\mathcal{L}$ is the integrated luminosity, $(1+\delta) \equiv \kappa$ is the ISR correction factor (see section~\ref{sec::ISRCORRECTION}), $\frac{1}{|1-\Pi(s)|^{2}}$ is the vacuum polarization correction factor (with values in the range $1.05$--$1.06$  from Ref.~\cite{FJegerlehner2017}), $\epsilon$ the efficiency from corresponding signal Monte Carlo after selection criteria, and $\mathcal{B}$ is the combined branching ratio of $\mathcal{B}({\chi_{cJ} \rightarrow \gamma J/\psi})$ and $\mathcal{B}({J/\psi \rightarrow \ell^{+} \ell^{-}})$.
The systematic uncertainties are taken into account by assuming a Gaussian-shaped uncertainty on the  
efficiency with a width equal to the total systematic uncertainty.

\begin{figure}[!htbp]
\centering
\includegraphics[width=8.6cm]{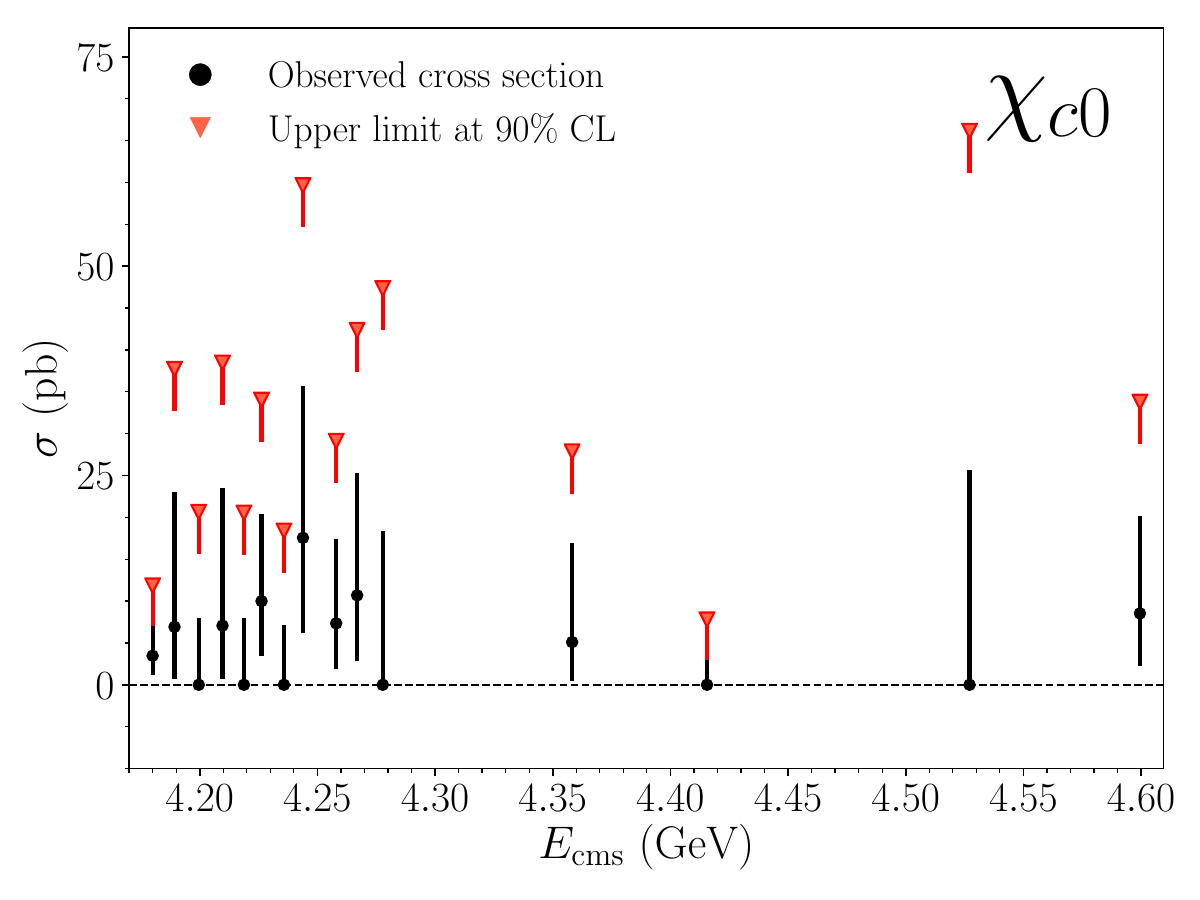}
\includegraphics[width=8.6cm]{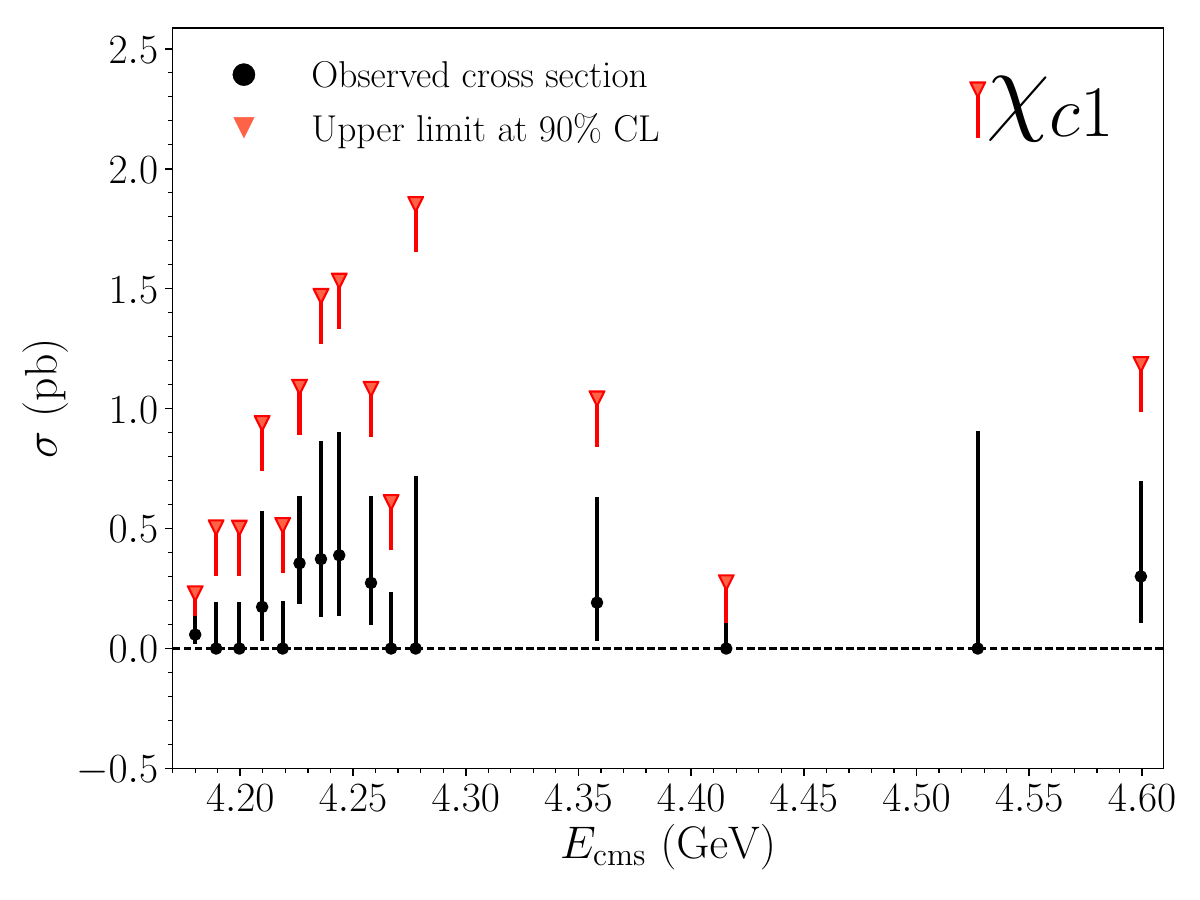}
\includegraphics[width=8.6cm]{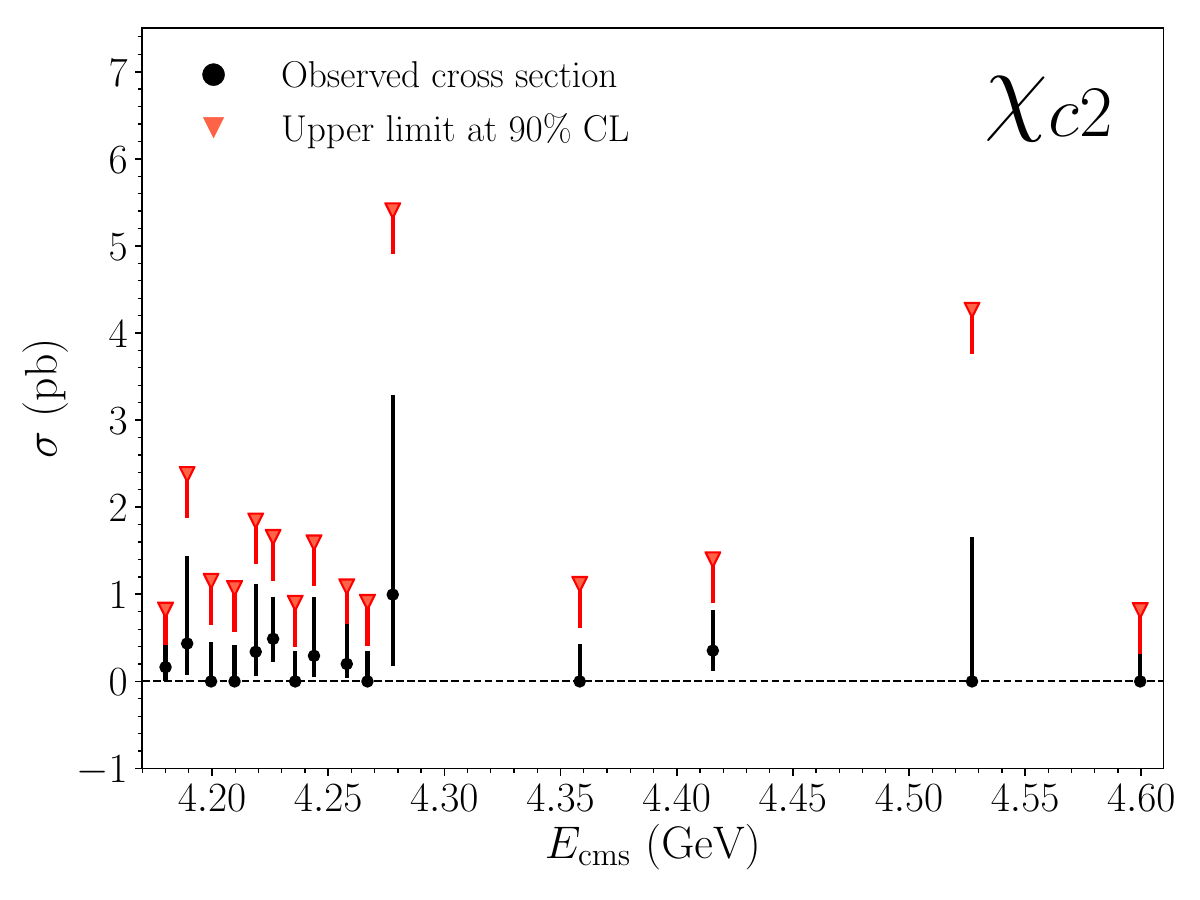}
\caption{ \label{figure::results}Cross section (black) and corresponding upper limit (red) for the reaction 
channels $e^{+}e^{-}\rightarrow \chi _{cJ}\pi^{+}\pi^{-}$ versus the center-of-mass energy E$_{\text{cms}}$.}
\end{figure}

The measured cross sections and the corresponding upper limits at the 90\% confidence level are
summarized in \cref{table::results0,table::results1,table::results2} and in Fig.~\ref{figure::results}.
The quoted statistical significance is based on the binomial assumption  $Z_{\text{Bi}}$, taken from Cousins {\it et al.}~\cite{COUSINS2008480} and does not include any systematic uncertainties.
With the exception of the channel $e^{+}e^{-}\rightarrow\pi^{+}\pi^{-}\chi _{c1}$, the measured cross sections show no significant variation with center-of-mass energy. 
It should be noted that the upper limits for $e^{+}e^{-}\rightarrow\pi^{+}\pi^{-}\chi _{c0}$ 
are less restrictive than those for the other two modes on account of the small branching 
ratio of $\chi _{c0}\rightarrow\gamma J/\psi$.
Since no convincing $\chi _{cJ}\pi^{+}\pi^{-}$ signal is seen, the quoted upper limits can also be 
considered as upper limits on the reaction proceeding through a hypothetical $Z_{c}(4050)^{\pm}$ particle.

\section{\label{sec::SUMMARY}Summary}
We have performed a search for the process $e^{+}e^{-}\rightarrow \chi _{cJ}\pi^{+}\pi^{-}$, $\chi _{cJ}\rightarrow\gamma  J/\psi$, $J/\psi\rightarrow (e^{+}e^{-}/\mu^{+}\mu^{-})$, at center-of-mass energies ranging from 4.18\,GeV to 4.60\,GeV. No significant signal has been observed, despite the hint of an slight enhancement for $\pi^{+}\pi^{-}\chi _{c1}$  at center-of-mass energies between 4.18\,GeV and 4.26\,GeV. Thus, we set upper limits at the 90\% CL for the three studied reaction channels for $J=0,1,2$. 
Since no signal is observed also no charmonium-like structure in the invariant mass of the $\chi _{cJ} \pi^{\pm}$ subsystem can be seen. So the upper limits of the reaction channels $\chi _{cJ}\pi^{+}\pi^{-}$ also apply for the case with an intermediate structure.

	\section*{Acknowledgments}
	The BESIII collaboration thanks the staff of BEPCII and the IHEP computing center for their strong support. This work is supported in part by National Key Basic Research Program of China under Contract No. 2015CB856700; National Natural Science Foundation of China (NSFC) under Contracts Nos. 11625523, 11635010, 11735014, 11822506, 11835012, 11935015, 11935016, 11935018, 11961141012; the Chinese Academy of Sciences (CAS) Large-Scale Scientific Facility Program; Joint Large-Scale Scientific Facility Funds of the NSFC and CAS under Contracts Nos. U1732263, U1832207; CAS Key Research Program of Frontier Sciences under Contracts Nos. QYZDJ-SSW-SLH003, QYZDJ-SSW-SLH040; 100 Talents Program of CAS; INPAC and Shanghai Key Laboratory for Particle Physics and Cosmology; ERC under Contract No. 758462; German Research Foundation DFG under Contracts Nos. Collaborative Research Center CRC 1044, FOR 2359; Istituto Nazionale di Fisica Nucleare, Italy; Ministry of Development of Turkey under Contract No. DPT2006K-120470; National Science and Technology fund; STFC (United Kingdom); The Knut and Alice Wallenberg Foundation (Sweden) under Contract No. 2016.0157; The Royal Society, UK under Contracts Nos. DH140054, DH160214; The Swedish Research Council; U. S. Department of Energy under Contracts Nos. DE-FG02-05ER41374, DE-SC-0012069; Olle Engkvist Foundation under Contract No. 200-0605

\FloatBarrier

\bibliography{Bibliography}

\end{document}